\documentclass[aps]{revtex4}
\begin{document}
\title{Editorial}
%
%
\date{\today}

\maketitle In the last ten years, an intense activity has been
modifying and improving our understanding of contemporary
Statistical Mechanics and Thermodynamics. After the success of
NEXT2001, held in Sardinia
in September 2001, where a large variety of systems whose
description appears to require the extension of the
Boltzmann-Gibbs formalism were discussed \cite{next2001},
this second conference on "News and expectations in
thermostatistics" NEXT2003 has brought together and synthesized the
latest work in a field where theory has reached a critical
turning point and several still unclear points have acquired sound
fundamental justifications. In addition to a critical assessment
of the results achieved, this conference has helped
cross-fertilization of interdisciplinary and new applications and
an unbiased discussion.

During the Conference about
one third of the presentations dealt with fundamentals and
theoretical methods of modern statistical mechanics and thermodynamics.
All the other presentations focused on applications of statistical mechanics
to specific fields: Hamiltonian
systems, chaos, fluctuations, stochastic systems, time series, models of
complex and quantum systems, gravitation and high energy physics.

As these proceedings together with those of the previous
edition of the conference \cite{next2001} give a good
overview of the large activity and variety of results in the field,
we devote
the editorial to a historical outline of the development of some
mathematical functions that play an important role in the current
theoretical foundations of the field.

In 1779, three years before his death, Euler studied the
properties of a series first introduced by Lambert in 1758
in order to
express roots of the trinomial algebraic equation $x=q+x^m$. In
his paper \emph{ De serie Lambertina plurimisque eius insignibus
proprietatibus} \cite{E} Euler considered the \emph{aequatio
trinomialis} in the symmetrical form
$x^{\alpha}-x^{\beta}=(\alpha-\beta)v\,x^{\alpha+\beta}$. Here he
introduced the function
\begin{eqnarray}
v=\frac{x^{-\beta}-x^{-\alpha}}{\alpha-\beta}  \label{1}
\end{eqnarray}
and its derivative. In particular he illustrated the special case
$\beta=0$ and the limiting case $\beta=\alpha$.

When discussing $\beta=0$, the function defined in Eq. (\ref{1}) becomes:
\begin{eqnarray}
v=\frac{1- x^{-\alpha}}{\alpha} \ , \label{2}
\end{eqnarray}
and Euler found its inverse:
\begin{eqnarray}
x=(1-\alpha \,v)^{-\frac{1}{\alpha}} \ ; \label{3}
\end{eqnarray}
while,
when studying the case $\beta=\alpha$, he put $\alpha =
\beta + \omega$ with $\omega\to 0$ ($\omega$ \emph{infinite
parvo}) to avoid an indeterminate form of the equation and
explicitly uses (page 353 of Ref. \cite{E}) the limit: \emph{
Constat autem evanescente $\omega$ esse}
\begin{eqnarray}
\frac{x^{\omega}-1}{\omega}=lx \ , \label{4}
\end{eqnarray}
where symbol $lx$ indicates the natural logarithm $\ln (x)$.

Euler used (\ref{1}),(\ref{2}) and (\ref{3}), and the limit
(\ref{4}) as mathematical tools to approach a specific
mathematical problem: after 130 years these same tools reappeared in
mathematical statistics and, 60 years later, in information
theory and statistical physics.

Indeed in 1908 W. S. Gosset, under the pen name \emph{Student},
published \cite{SG} the distribution function:
\begin{eqnarray}
y= A_n \,(1+z^2)^{-\frac{1}{2}n} \ , \label{5}
\end{eqnarray}
where the free parameter $n$ is an integer. Guided by the distribution (\ref{5}) and
by the subsequent work in mathematical statistics, in 1968 V. M. Vasyliunas \cite{V}
introduced the phenomenological distribution:
\begin{eqnarray}
y= C_{\kappa} \left(1+\frac{v^2}{\kappa
\,\theta^2}\right)^{-(\kappa + 1)}   \label{6}
\end{eqnarray}
to reproduce cosmic ray energy distributions:  Eq. (\ref{6}) goes
smoothly to  the Maxwellian distribution
when the continuous parameter $\kappa$ of Vasyliunas
approaches infinity. Nowadays
distribution (\ref{6}) is widely used in
plasma physics where it is  known as
$\kappa$-distribution.

Before discussing the successive applications of
Euler's Eqs.  (\ref{1})-(\ref{4}) in
statistical mechanics and information theory,  we
should recall a few points concerning entropy.
A generalized trace-form entropy can be written as:
\begin{eqnarray}
{\cal S}(p)=\sum_{i}p_i \,f(1/p_i) \ , \label{7}
\end{eqnarray}
where $f(x)$ is an arbitrary function whose properties have been
studied by Csiszar \cite{C}. This function can be viewed
as a generalization of the logarithm since, when $f(x)=\ln x$, the
above generalized entropy reduces to the Boltzmann-Gibbs-Shannon
entropy. After the publication in 1948 of Shannon's paper
\cite{S}, other entropies have been proposed within information
theory: many of them are of the form (\ref{7}) for specific
choices of $f(x)$.

In information theory and mathematical statistics parameter
interpretation is less important and Boltzmann-Gibbs-Shannon
entropy has been generalized in about thirty different ways adding
one or two parameters. Distribution functions
are even more diversified: a handbook of generalized special
functions for statistical and physical sciences by Mathai~\cite{MA} lists
more than 130 different statistical distributions.

In 1967 Harvda and Charvat proposed a one-parameter
generalization of Shannon information entropy of the form
(\ref{7}) with a choice of $f(x)$ which is basically Eq.
(\ref{2}).

In  1975, in the context of information theory,
Mittal \cite{M} and Sharma and Taneja \cite{ST} proposed
a further generalization of the entropy of Harvda and Charvat,
 using as $f(x)$ the
two-paramter dependence (\ref{1}). This entropy has not been as
successful as the one of Harvda-Charvat.

In 1988 Tsallis, motivated by multi fractal scaling, first
postulated the physical relevance of a one-parameter
generalization of the entropy \cite{T}, whose form is
equivalent to that of Harvda and Charvat. This seminal paper
\emph{opens the door to the generalization of standard
thermodynamics} and of the Boltzmann-Gibbs statistical mechanics.

In Tsallis' theory, which is often referred to as non-extensive
statistical mechanics, generalizations of the logarithm and of
the exponential have an important role:
\begin{equation}
\ln_q(x)=\frac{x^{1-q}-1}{1-q} \ , \quad\quad
\exp_q(x)=\left[1+(1-q)x\right]_+^{\,\,1/(1-q)}\ , \label{8}
\end{equation}
where $[x]_+ \equiv \max(x,0)$. These two functions, which
coincide with Eqs. (\ref{2}) and (\ref{3}), have appeared for the first
time in the same coherent physical context linked by the maximum
entropy principle.

Tsallis' entropy:
\begin{eqnarray}
{\cal S}_q(p)=\sum_{i}p_i \,\ln_q(1/p_i) \ , \label{9}
\end{eqnarray}
and the related distribution function expressed in terms of
$\exp_q(x)$ represent the foundation of the non-extensive
statistical mechanics: the reader is referred to the large number
of papers that deal with this theory and its applications, both
in this volume and in the previous one \cite{next2001}, for a
more complete discussion.

In 1997 Abe generalized statistical mechanics
in a new direction  \cite{A} within the framework of quantum
groups. Soon afterwards Borges and Roditi \cite{BR}
found a unified formulation for Tsallis' and Abe's entropy. They
proposed a two-parameter entropy that happened to be that of
Mittal-Sharma-Taneja. In fact Borges and Roditi have constructed with
this entropy a generalized statistical mechanics that yields a
two-parameter distribution which decays as a power law. Only for
special choices of the parameters, such as the choice that yields
Tsallis' statistical mechanics, this distribution can be obtained
explicitly without resorting to numerical evaluation.

The distribution function corresponding to Mittal-Sharma-Taneja entropy
decays as a power law: one of the two parameters is immediately related to
this asymptotic behavior, while the second one is of more difficult
interpretation. This fact might explain why this two-parameter entropy has
not been used by the physics community.
A simple expression for the entropy and the distribution
function and a minimal number of
parameters are important features of a physical theory.

Very recently \cite{K} the requirement that the generalized
logarithm  verifies $f(1/x)=-f(x)$ has produced a new family of
trace-form entropies  where in Eq. (\ref{7})
$f(x)=g\left((x^{\kappa}-x^{-\kappa})/(2\kappa)\right)  $ with
$g(x)$ an arbitrary, odd, and increasing function. For the
special case $g(x)=x$ one obtains the following expressions for
the generalized logarithm and its inverse function, \emph{i.e.},
the generalized exponential:
\begin{equation}
\ln_{{\scriptscriptstyle \{}\kappa { \scriptscriptstyle
\}}}(x)=\frac{x^{\kappa}-x^{-\kappa}}{2\kappa}\, , \quad\quad
\exp_{{\scriptscriptstyle \{}\kappa { \scriptscriptstyle
\}}}(x)=\left(\sqrt{1+\kappa^2x^2}+\kappa x \right)^{1/\kappa}
 \quad . \label{11}
\end{equation}
Using the one-parameter generalized entropy
\begin{eqnarray}
{\cal S}_{\kappa}(p)=\sum_{i}p_i \,\ln_{{\scriptscriptstyle
\{}\kappa { \scriptscriptstyle \}}}(1/p_i)=-\sum_{i}p_i
\,\ln_{{\scriptscriptstyle \{}\kappa { \scriptscriptstyle
\}}}(p_i) \ ,  \label{12}
\end{eqnarray}
the maximum entropy principle yields the
one-parameter distribution function:
\begin{equation}
 p_i=\alpha\exp_{{\scriptscriptstyle \{}\kappa {
\scriptscriptstyle \}}} \left( -\frac{E_i-\mu}{\lambda T} \right)
\quad ,\label{12bis}
\end{equation}
which decays asymptotically as a power law  and reproduces the
Maxwell-Boltzmann distribution in the limit $\kappa\to 0$. The
structure of the ensuing generalized statistical mechanics has
striking similarity with that of special relativity suggesting
that it might be relevant for systems where information
propagates with finite speed, \emph{e.g.}, relativistic particle
systems \cite{K}.

As useful mathematical tools, one could find many two-parameter
generalizations of the logarithm that reproduce as special cases
the $q-$logarithm or the $\kappa-$logarithm.
A simple two-parameter
logarithm, which has the important property that its inverse
exists in terms of elementary functions, is the scaled
$\kappa-$logarithm
\begin{eqnarray}
\ln_{{\scriptscriptstyle \{}\kappa , \varsigma { \scriptscriptstyle
\}}}(x) \equiv \frac{2}{\varsigma^{\kappa}+\varsigma^{-\kappa}} \, \left [ \,
\,\ln _{{\scriptscriptstyle \{}\kappa { \scriptscriptstyle \}}}(\varsigma
x)-\ln_{{\scriptscriptstyle \{}\kappa { \scriptscriptstyle
\}}}(\varsigma)\,\right ] \ , \label{13}
\end{eqnarray}
where $\varsigma$ is the scaling parameter. Note that this
two-parameter generalization of the logarithm, $\ln (x)=
\ln_{{\scriptscriptstyle \{}0 ,  \varsigma\, { \scriptscriptstyle \}}}(x)
$, contains as special cases both
the $q-$logarithm, $\ln_q(x)
=\ln_{{\scriptscriptstyle \{}q-1, 0 { \scriptscriptstyle \}}}(x)$
and the $\kappa-$logarithm,
$\ln_{{\scriptscriptstyle \{}\kappa { \scriptscriptstyle \}}}(x)
=\ln_{{\scriptscriptstyle \{}\kappa , 1 { \scriptscriptstyle
\}}}(x)$.

A second example
of two-parameter generalization of the logarithm is given
by Euler's function (1), which includes both the q-logarithm
and the $\kappa$-logarithm.
This latter special case, the
$\kappa$-logarithm, has not been discussed either by Euler, or
by Mittal \cite{M}, or by Sharma-Taneja \cite{ST}, or by Borges-Roditi
\cite{BR}, when using (\ref{1}). Nor has
the $\kappa$-exponential, in spite of its simplicity,
attracted the attention of researchers or been included in
the large compilation of statistical distributions by Mathai [13].

We observe that, starting from Euler's work, the fundamental
limit (\ref{4}), on which the definition of natural logarithm and
the constant $e$ itself are based,
has been used by generations of scientists
and students. After about two centuries, the founders of
generalized information theory and non-extensive statistical
mechanics resorted to this same formula to generalize the entropy of
Boltzmann-Gibbs-Shannon paying an implicit tribute to Euler, one of the
most prolific mathematician in history.

We close this editorial with our deepest thanks to the Chairmen,
Eddy Cohen, Piero Quarati, the proponent of NEXT2001 and NEX2003, and
Constantino Tsallis; to the distinguished scientists of
the International Advisory Committee; to Alberto Devoto,
Giuseppe Mezzorani, Andrea Rapisarda, Lamberto  Rondoni, and
Stefano Ruffo, which took care with us of the local organization;
to our sponsors ``Regione Sardegna'', ``Ministero
dell'Istruzione, dell'Universit\`a e della Ricerca'' (MIUR),
``Universit\`a di Cagliari'', ``Universit\`a di Catania'',
``Politecnico di Torino'', ``Istituto Nazionale di Fisica
Nucleare'' (INFN), ``Istituto Nazionale di Fisica della Materia''
(INFM), ``Associazione Sviluppo Scientifico e Tecnologico
Piemonte'' (ASP), and ``Ente Sardo Industrie Turistiche'' (ESIT).
In addition we congratulate Eddy Cohen and Gene Stanley for
the 2004 Boltzmann award for their outstanding
achievements in Statistical Physics.

\vskip 1cm
Giorgio Kaniadakis \\
Dipartimento di Fisica, INFM \\
Politecnico di Torino \\
Corso Duca degli Abruzzi, 24 \\
I-10129 Torino, Italy \\
Email address: {\tt giorgio.kaniadakis@polito.it}
\vskip 1cm
Marcello Lissia \\
INFN and Physics Dept. \\
Universit\`a di Cagliari \\
Cittadella Universitaria \\
I-09042 Monserrato (CA), Italy \\
Email address: {\tt marcello.lissia@ca.infn.it}

\end{document}